\documentstyle[prl,aps,twocolumn]{revtex}
\catcode`\@=11

\def\maketitle2{\par % Uses \twocolumn[\@maketitle2].
\begingroup
\let\cite\@bylinecite
\def\thefootnote{\fnsymbol{footnote}}%
\twocolumn[\@maketitle2\vskip2pc]%
\thispagestyle{plain}\@thanks
\endgroup
\def\thefootnote{\arabic{footnote}}%
\setcounter{footnote}{0}%
\let\maketitle2\relax \let\@maketitle2\relax
\let\@thanks\relax \let\@authoraddress\relax \let\@title\relax
\let\@date\relax \let\thanks\relax \let\@abstract\relax 
\let\@pacs\relax}

\def\abstract#1{\gdef\@abstract{{\par % Store abstract text. 
\bgroup
\ifdim\prevdepth=-1000pt \prevdepth0pt\fi
\hsize\columnwidth
\dimen0=-\prevdepth \advance\dimen0 by17.5pt \nointerlineskip
\small\vrule width 0pt height\dimen0 \relax}{~~}#1\egroup}}

\def\pacs#1{\gdef\@pacs{{\par % Store PACS numbers as \@pacs.
\bgroup
\hsize\columnwidth \parindent0pt
\ifdim\prevdepth=-1000pt \prevdepth0pt\fi
\dimen0=-\prevdepth \advance\dimen0 by20pt\nointerlineskip
\egroup} PACS numbers:~#1}}

\def\@maketitle2{% Puts \@abstract and \@pacs in a {list}.
\@preprint
\@title
\ifdim\prevdepth=-1000pt \prevdepth0pt\fi
\@authoraddress
\@date
\begin{list}{}{\leftmargin=0.10753\textwidth \rightmargin=\leftmargin
\itemsep=1pc\partopsep=-1pc}
\item\@abstract
\item\@pacs
\end{list}
}

\catcode`\@=12

\begin{document}

\title{Indication, from Pioneer 10/11, Galileo, and Ulysses Data, \\
of an Apparent Anomalous, Weak, Long-Range Acceleration$^+$}
	
\author{John D. Anderson,$^a$ Philip A. Laing,$^b$ 
Eunice L. Lau,$^c$ Anthony S. Liu,$^d$ \\
Michael Martin Nieto,$^e$ and Slava G. Turyshev$^f$} 

\address{$^{a,c,f}$Jet Propulsion Laboratory, California Institute of 
Technology, Pasadena, CA 91109}   
\address{$^b$ The Aerospace Corporation, 2350 E. El Segundo Blvd., 
El Segundo, CA 90245-4691} 
\address{$^d$ Astrodynamic Sciences, 2393 Silver Ridge Ave., 
Los Angeles, CA 90039} 
\address{$^e$ Theoretical Division (MS-B285), 
Los Alamos National Laboratory,
University of California, Los Alamos, NM 87545}
%***********************************************
\date{\today}
%**************************************************
\abstract
{Radio metric data from the Pioneer 10/11, Galileo, and 
Ulysses spacecraft   
indicate an apparent anomalous, constant, acceleration acting on the  
spacecraft with a magnitude  $\sim 8.5
\times 10^{-8}$ cm/s$^2$, directed towards the Sun.  Two independent
codes and physical strategies have been used to  analyze the data.   
A number of potential causes have been ruled out.  
We discuss future kinematic tests and possible origins of the
signal.}


%******************************FOR LANL*************
%\maketitle                %****REVTEX***************
%***************************************************
\maketitle2              %****MACROS***************
\narrowtext              %*************************
%***************************************************

Exploration of the outer planets began with the launch of 
Pioneer 10 on 2 March 1972,  \cite{science}. (Pioneer 11 followed on 5
April 1973.)  After Jupiter and  (for Pioneer 11) Saturn encounters, the
two spacecraft followed hyperbolic orbits near the plane of the ecliptic to 
opposite sides of the solar system.  Although Pioneer 10 is still
transmitting, its mission officially ended on 31 March 1997 when it was at
the distance of 67 Astronomical Units (AU) from the Sun.   Pioneer 11's
radio system failed and coherent Doppler signals were last received on 1
October 1990,  when the spacecraft was 30 AU away from the Sun.  

The Pioneer spacecraft are excellent for dynamical astronomy studies. 
Due to their spin-stabilization and their great distances, 
a minimum number of Earth-attitude reorientation maneuvers are required. 
This permits precise acceleration estimations,   to the level
of $10^{-10}$ cm/s$^2$ (averaged  over 5 days). Contrariwise, a
Voyager-type spacecraft is not well suited for a precise
celestial mechanics experiment as its numerous  attitude-control
maneuvers overwhelm any  small external acceleration.

%***************************************************************

To obtain the S-band Doppler data from the Pioneer spacecraft, NASA used
the Jet Propulsion Laboratory's (JPL) Deep Space Network (DSN).   
This data was used in 
the two analyses described below  to determine
Pioneer's initial position, velocity and  the magnitudes of the
orientation maneuvers. The analyses were modelled to include the
effects of planetary perturbations, radiation pressure, the interplanetary 
media, general relativity, and  bias and drift in the range and Doppler. 
Planetary coordinates and the solar system masses 
were obtained using JPL's Export Planetary  Ephemeris DE200.  Both
analyses calculated Earth's polar motion and its  non-uniform rotation
using the International Earth  Rotation Service.

%**********************************************************

Beginning in 1980, when at 20 AU the solar radiation pressure 
acceleration had decreased to  
$< 5 \times 10^{-8}$ cm/s$^2$  \cite{jpl},  JPL's 
Orbit Determination Program (ODP) analysis of unmodelled 
accelerations (at first with the faster-moving Pioneer 10) 
found that the biggest systematic error in the acceleration residuals
is a constant bias of $a_P \sim (8\pm 3) \times 10^{-8}$ cm/s$^2$, 
directed {\it toward} the Sun \cite{bled}, 
to within the accuracy of the Pioneers' antennae.   As possible 
``perturbative forces" to explain this bias, we considered gravity 
from the Kuiper belt, gravity from the galaxy,
spacecraft ``gas leaks,'' errors in the planetary ephemeris, and
errors in the accepted values of the
Earth's orientation, precession,  and nutation. 
None of these ``forces"  explained the apparent acceleration. 
Some were three orders of magnitude or more too small.  

%**********************************************************
Non-gravitational effects, such as solar radiation pressure
and precessional attitude-control maneuvers, make small contributions to
the apparent acceleration we have observed.   The solar radiation
pressure decreases as $r^{-2}$.   As 
previously indicated for the Pioneers, 
at distances $>$10-15 AU it produces an  acceleration that is 
much less than $8\times 10^{-8}$ cm/s$^2$, directed 
{\it away} from the Sun. (The solar wind is
roughly a factor of 100 smaller than this.) 

A possible systematic explanation of the residuals is  non-isotropic
thermal radiation.   
Pu$^{238}$ (half life of 87.74 years) 
radioactive thermal generators (RTGs) power the Pioneers. 
At launch the RTGs delivered 160 W of
electric power.  Power has decreased approximately linearly ever 
since.  By 1997 a little less than 80 W were available. 
The excess power and the heat generated by the plutonium  
has been thermally radiated into space.   The power needed
for this to  explain $a_P$  is $\sim 85$ W. There
is almost that much available, but presumably the radiation was
approximately isotropic.  Further, if it were not, and was the cause of
$a_P$,  this acceleration would have decreased with
time. After 1980, no such   
(linearly decreasing) acceleration was observed.  
%********************************************************
Another radiation source is the Pioneer radio beam. The power 
emitted from the antenna is 8 W.  This implies a bias maximum of 
less than 9\% of $a_P$, and in the {\it opposite}  direction.  
(The influence of the bias is being investigated.)
%********************************************************

We conclude, from the JPL-ODP analysis,  that there is an unmodelled 
acceleration, $a_P$, towards the Sun  of $(8.09\pm0.20)\times 10^{-8}$ 
cm/s$^2$ for Pioneer 10 and of $(8.56\pm 0.15) \times 10^{-8}$ cm/s$^2$  
for Pioneer 11.  The error is determined by use of a 
five-day batch sequential filter with radial acceleration as a
stochastic parameter subject to white Gaussian noise ($\sim$ 500
independent five-day samples of radial acceleration) \cite{tap,error}.
No magnitude variation of $a_P$ with distance was found, within a  
sensitivity of 2 $\times$ 10$^{-8}$ cm/s$^2$ over a 
range of 40 to 60 AU.
%****************************************************************

Continuing our search for an explanation, we considered the 
possibilities i) that the Pioneer 10/11 spacecraft had internal 
systematic properties, undiscovered because they are of identical
design,  and ii) that the acceleration was due to some  not-understood
viscous drag force (proportional to the approximately  constant velocity
of the Pioneers). Both these possibilities
could be investigated by studying  spin-stabilized craft whose spin axes 
are not directed towards the Sun, and whose orbital velocity vectors 
are far from being radially directed.  

Two candidates were Galileo in its  Earth-Jupiter mission phase and 
Ulysses  in Jupiter-perihelion  cruise out of the plane of the 
ecliptic.  As well as Doppler,
these spacecraft also yielded a considerable quantity of range data.  
Ranging data are generated by cross correlating a phase modulated 
signal with a ground duplicate and noting the time delay. 
Thus, the ranging data are 
independent of the Doppler data, which represent a frequency shift 
of the radio carrier wave without modulation. (For example, solar plasma 
introduces a group delay in the ranging data
but a phase advance in the Doppler data.)  Ranging data can be used to 
distinguish an actual range change from a fictitious one caused by a 
frequency error.  

A quick look at Galileo showed it was impossible to separate 
the solar radiation effect from the anomalous constant acceleration with 
the limited data analyzed  
(241 days from 8 January 1994 to 6 September 1994) \cite{johnson}.

However, an analysis of the radiation pressure on Ulysses 
in its out-of-the-ecliptic journey, from 5.4 AU near Jupiter
in February 1992 to the perihelion at 1.3 AU in February 1995, 
found a varying profile with distance \cite{uly}. 
The orbit solution requires a periodic updating of the
solar radiation pressure. The radio Doppler and
ranging data can be fit to the
noise level with a time-varying solar constant in the fitting model
\cite{mcelrath}.  The inferred solar constant is about 40 percent
larger at perihelion (1.3 AU) than at Jupiter (5.2 AU), a physical
impossibility. By interpreting this time variation 
as a true $r^{-2}$ solar pressure plus a constant radial
acceleration, we conclude that Ulysses was subjected to an
unmodelled acceleration  towards the Sun of  
(12 $\pm$ 3) $\times 10^{-8}$ cm/s$^{2}$.

%********************************************************************

With no explanation of this data in hand, our attention focused on
the possibility that there was some error in  JPL's ODP.  To investigate
this, an independent analysis of the raw data using The
Aerospace Corporation's Compact High Accuracy Satellite Motion Program
(CHASMP), which was developed independently
of JPL's ODP, was performed. Although by necessity, both programs use
the same physical principles,  planetary ephemeris, and timing and polar
motion inputs, the algorithms are  otherwise quite different. If there
were an error in either program, they would not agree. (Common program
elements continue to be investigated.)

The CHASMP analysis of Pioneer 10 data
also showed an unmodelled acceleration 
in a direction along the radial toward the  Sun \cite{aero}. 
The value is $(8.65 \pm 0.03) \times 10^{-8}$ cm/s$^{2}$, 
agreeing with JPL's result. 
The smaller error here is because the CHASMP analysis 
used a batch least-squares fit over the whole orbit \cite{tap}, not looking 
for a variation of the magnitude of $a_P$ with distance.

Without using the apparent acceleration, 
CHASMP shows a steady frequency drift of about
$-6 \times 10^{-9}$ Hz/s, or 1.5 Hz over 8 years 
(one-way only). This equates to
a  clock acceleration, $-a_t$, of $-2.8\times 10^{-18} s/s^{2} $. 
The identity with $a_P$ is $a_P \equiv a_t c$. 
The drift in
the Doppler residuals (observed minus computed data) is 
seen in Figure 1.  It is clear, definite, and  cannot be removed without
either the added acceleration, $a_P$,  
or the inclusion in the data itself of a frequency drift, 
i.e., a ``clock acceleration'' $a_t$.  

If there were a systematic drift in the atomic clocks of the DSN  or in
the time-reference standard signals, this would appear  
like a non-uniformity of time; i.e., all clocks would be changing with a 
constant acceleration. We have not yet been able to rule out this
possibility.  Elements common to the Doppler and range 
tracking systems (e.g., DSN station clocks) need to be investigated.  For 
example, how and to what accuracy are the clocks at different DSN stations 
tied to each other and to external national standards? Are there differences 
in the orbital fits when different stations' data are analyzed separately? 

%*******************************************************************

Aerospace's analysis of Galileo data covered the same arc as JPL and a 
second arc from 2 December 1992 to 24 March 1993.  
Doppler data from the first arc
resulted in a determination for $a_P$ of  
$\sim (8 \pm 3) \times 10^{-8}$ cm/s$^2$,  
a value similar to that from Pioneer 10.  But the correlation
with solar pressure was so high (.99) that it is impossible to decide
whether solar pressure is a contributing factor. 
[Galileo is less sensitive to both the $a_P$- and 
$a_t$-model effects than the Pioneers. 
Pioneers have a smaller solar pressure and a longer light travel time. 
Sensitivity to a clock acceleration is 
proportional to the light travel time squared.]
The second arc was 113 days long, starting  
six days prior to the second Earth encounter.
This solution  was also too highly
correlated with solar pressure, and the data analysis was 
complicated by many mid-course maneuvers.
The maneuver uncertainties 
were so great, a standard null result could not be ruled out.

However, there was an additional result from this second 
arc.  This arc was chosen for study because there was ranging data. 
The two-way range change and time integrated 
Doppler are consistent to $\sim 4$ m over 
a time interval of one day. 
This is strong (but not conclusive) evidence 
that the apparent acceleration is
not the result of hardware problems at the tracking stations.

%*********************************************************************

With these added discoveries, what other possible origins for the 
signal come to mind?  

One can speculate that there is some unknown interaction of the radio
signals with the solar wind.  An experimental
answer  could be given with two different transmission frequencies.  
Although the main communication link on the Ulysses mission 
is S-up/X-down mode, a small fraction of the data is S-up/S-down.  We
plan to utilize this option in  further analysis.
%********************************************************************** 

If no normal explanation for the residuals is found, further  
tests of the effect are needed.   The weakening Pioneer 10
signal can still be reacquired for a short time.  (The NASA/Ames
Lunar Prospector Team has intermittently done this for training
purposes, producing high quality data.)  
Further Ulysses data would also help. 

The Pluto Express mission could provide an excellent opportunity
for high-quality data from very deep space, especially if 
optical tracking is used.   A similar opportunity may exist, out
of the plane of the ecliptic, from the proposed Solar Probe mission. 
Under consideration is  a low-mass module to be ejected 
during solar fly by.  

%*******************************************************************

With all the above, it is interesting to speculate on the 
possibility that the origin of the anomalous signal 
is new physics \cite{photon}.  This is true 
even though the  probability is that some ``standard physics"
or some as-yet-unknown systematic will be found to explain this  
``acceleration."  This probability is of interest in itself, given 
that we have found no plausible explanation so far.

The paradigm is obvious. 
``Is it dark matter or a modification of gravity?"  
Unfortunately, neither easily works. 

If the cause is dark matter, it is hard to understand.   The
spherically-symmetric distribution of matter, 
$\rho \sim r^{-1}$, produces a constant acceleration {\it inside} 
the distribution.  For this to cause $a_P$, even  only out to
50 AU, would require the total dark matter to be  $>~3 \times
10^{-4}  M_\odot$. But this  is in conflict with the
accuracy of the ephemeris, which allows only  of order a few times
$10^{-6} M_\odot$ of dark matter even within the  orbit of Uranus
\cite{ephem}. (A 3-cloud neutrino model also did not solve 
the problem \cite{jgscold}.)

%******************************************************************

Contrariwise, the most commonly studied 
possible modification of gravity (at various scales) is an 
added Yukawa force \cite{physrep}. Then the 
gravitational potential is 
\begin{equation}
V(r) = -{GMm}[(1+\alpha)r]^{-1}[1 +\alpha\exp^{-r/\lambda}],
  \label{V}
\end{equation}
where $\alpha$ is the new coupling strength relative to
Newtonian gravity, and $\lambda$ is the new force's range.
Since the radial force is $F_r = -d_r V(r) =ma$,
the power series for the acceleration 
yields an inverse-square term, no inverse-$r$ term, then a constant
term.   Identifying this last term as the Pioneer acceleration yields 
\begin{equation}
a_P = -{a_1\alpha}[2(1+\alpha)]^{-1}
        [{r_1^2}/{\lambda^2}],   \label{solution}
\end{equation}
where $a_1$ is the Newtonian
acceleration at distance $r_1 =1$ AU.  (Out to 65 AU there
is no observational evidence of an $r$ term in the acceleration.) 
Eq. (\ref{solution}) is the solution curve; for example, 
$\alpha = -1 \times 10^{-3}$ for $\lambda = 200$ AU.  

It is also of interest to 
consider Milgrom's proposed modification of gravity \cite{mil}, 
where $a \propto 1/r^2$ for some constant $a_0 \ll a$ and 
$a \propto 1/r$ for $a_0 \gg a$.
Depending on the 
value of the Hubble constant, we find that $a_0 \approx a_P$.  

Of course, there are (fundamental and deep) theoretical problems if 
one has a new force of  the phenomenological types of those above.
Even so, the deep-space data piques our curiosity.  
However, these and other universal-gravitational explanations 
for the Pioneer effect come up against a hard experimental wall. 

%*********************************************************************

The anomalous acceleration  is too
large to have gone undetected in planetary orbits, particularly
for Earth and Mars.  NASA's Viking mission
provided radio-ranging measurements 
to an accuracy of about 12 m \cite{reasenberg,mg6}.
If a planet experiences a small, anomalous, radial acceleration,
$a_A$,  its orbital radius $r$ is
perturbed  by  
\begin{equation}
\Delta r =-{\it l}^6 a_A/(GM_\odot)^4 
         \rightarrow  - r [a_A/a_N] , 
  \label{deltar}
\end{equation}
where {\it l} is the orbital angular momentum per unit mass and $a_N$ is
the Newtonian acceleration at $r$. 
(The right value in Eq. (\ref{deltar}) holds in the circular orbit limit.)

\indent For  Earth and Mars, $\Delta r$ is about 
-21  km and -76 km.  However,
the Viking data determines the difference 
between the Mars and Earth orbital radii to about a
100 m accuracy, and  their sum 
to an accuracy of about 150 m.  The Pioneer effect is not seen.  

Further, a perturbation in $r$ produces a perturbation to the
orbital angular frequency of 
\begin{equation}
\Delta \omega =  2{\it l}a_A/(GM_\odot)
   \rightarrow 2 \dot{\theta} [a_A/a_N] .
\end{equation} 
The determination of the
synodic angular frequency $\omega_E - \omega_M$ is accurate to 7 parts
in 10$^{11}$, or to about 5 ms accuracy in synodic period. The
only parameter that could possibly mask the spacecraft-determined 
$a_R$ is  $(GM_\odot)$. But a large error here 
would cause inconsistencies with the overall planetary ephemeris
\cite{ephem,standish}. 

We conclude that the Viking ranging data limit any unmodelled radial
acceleration acting on Earth and Mars to no more than
$0.1 \times 10^{-8}$ cm/s$^2$. Consequently, if the anomalous radial
acceleration acting on spinning spacecraft is gravitational in origin, it
is {\it not} universal.  That is, it must affect bodies in the 1000 kg
range more than bodies of planetary size by a factor of 100 or more.  
This would be a strange violation of the Principle of Equivalence (PE) 
\cite{pe}. 
The fact an anomalous signal is not seen in the
analysis of the Viking Lander ranging data gives us added
confidence that the anomaly is not related to DSN hardware.
However, the Viking Lander data have not been analyzed by either
ODP or CHASMP, so we cannot make a similar claim regarding software
errors.

Similarly, the $\Delta \omega$ results rule out the universality of 
the $a_t$ time-acceleration model.  In the age of the universe, $T$,  
one would have $a_t T^2/2 \sim 0.7~T$.  
(Another unusual possibility is that there is some unknown 
non-kinematic effect causing a Doppler anomaly.)     

%***************************************************************
Clearly, more analysis, 
observation, and theoretical work are called for.  
Further details will appear elsewhere.   

We thank John E. Ekelund, Gene L. Goltz, William E. Kirhofer, Margaret
Medina, William L. Sjogren and S. Kuen Wong of JPL for their support in
obtaining and understanding  DSN Tracking Data. 
We also thank John W. Dyer, Alfred S. Goldhaber, 
Jack G. Hills, Irwin I. Shapiro, and Richard J. Terrile 
for many helpful conversations.  
This work was supported
by the Pioneer Project, NASA/Ames Research Center,
and was performed at the Jet Propulsion Laboratory, California Institute of
Technology, under contract with 
NASA. P.A.L. and A.S.L. acknowledge support by a grant from NASA
through the Ultraviolet, Visible, and Gravitational Astrophysics Program.
M.M.N. acknowledges support by the U.S. DOE.

%***********************************************************

%***********************************************************************
%************************************************************
\begin{figure}
\caption[Figure 1]{\small{
Two-way Doppler residuals (observed Doppler velocity minus model 
Doppler velocity) for Pioneer 10 in mm/s vs. time. 
Solar system gravity is represented by the Sun and the planetary
systems \cite{standish}. [If one adds one more
parameter to the model (a constant radial acceleration) the residuals are
distributed about zero Doppler velocity with a systematic variation 
$\sim$ 3.0 mm s$^{-1}$ on a time scale
$\sim$ 3 months.]  The outliers on the plot were rejected from the fit.
}}
\end{figure}
%****************************************************
%***************************************************


\begin{references}
\bibitem{error} All errors are from the
covariance matrices associated with the least--squares  
analysis \cite{tap}.
The assumed data errors are larger than the standard error on the 
post--fit residuals.  (For example, the Pioneer S--band Doppler error 
was set at 1 mm s$^{-1}$ at a Doppler integration time of 60 s, 
as opposed to a characteristic
chi-squared value of 0.3 mm s$^{-1}$.)  Thus, the quoted errors are 
realistic, not formal, representing an attempt to include systematics 
and reddening of the noise spectrum by solar plasma. Any spectral 
peaks in the postfit Pioneer Doppler residuals were not 
significant at a 90 percent confidence level.
\bibitem{pe}  The PE figure of merit is $a_P/a_N$.  This is worse 
than for lab experiments (comparing small objects) or for the
Nordtvedt Effect (large objects) \cite{will}.  It again emphasizes 
that Earth and Mars do not change positions due to $a_P$.
\end{references}
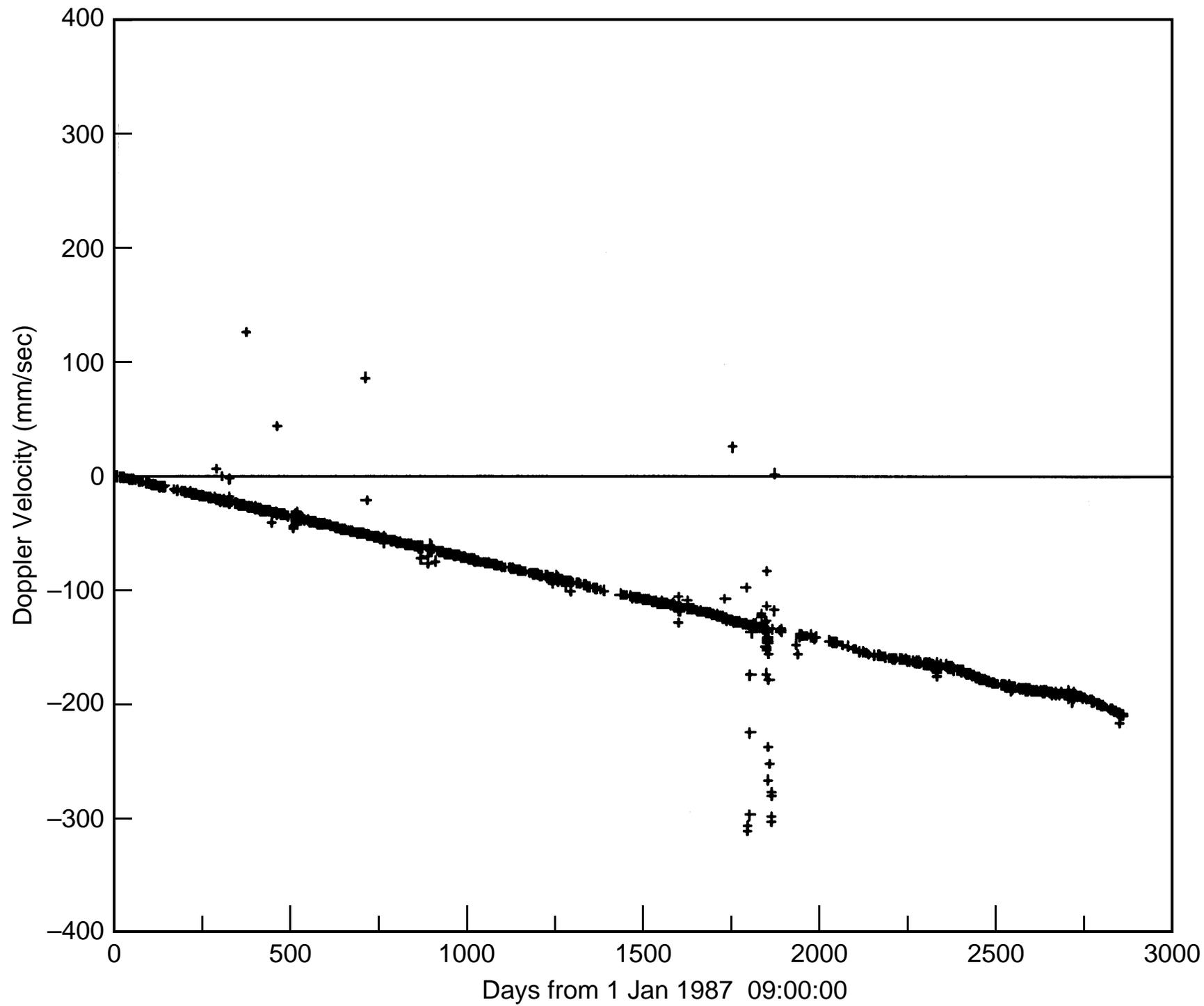
\end{document}